\documentclass[twocolumn, prd, nofootinbib, showpacs, amsmath, amssymb] {revtex4}

\usepackage{graphicx}

\begin{document}

\title{Cosmological implications of K\'arolyh\'azy uncertainty relation }

\author{ Michael~Maziashvili}
\email{maziashvili@hepi.edu.ge}
\affiliation{ Andronikashvili Institute of Physics, 6 Tamarashvili St., Tbilisi 0177, Georgia }

\begin{abstract}

K\'arolyh\'azy uncertainty relation, which can be viewed also as a relation
between UV and IR scales in the framework of an effective quantum field theory
satisfying a black hole entropy bound, strongly favors the existence of
dark energy with its observed value. Here we estimate the dynamics of dark
energy predicted by the K\'arolyh\'azy relation during the cosmological
evolution of the universe.

\end{abstract}





\maketitle

\section{Introduction}

From the inception of quantum mechanics the concept of measurement
(real or {\tt Gedanken}) has proved to be a fundamental notion for
revealing a genuine nature of physical reality. It is not without interest to
address from this standpoint a notion of background space-time. That is, to
ask how does the background space-time manifest itself in a way accessible to
us in light of quantum mechanics and general relativity. General relativity
treats space-time as a four-dimensional differentiable manifold with well
defined metric structure of Minkowskian signature. Physically it is easy to
understand that the introduction of a measuring device which undergoes quantum
fluctuations does not allow one to measure a background space-time with
unlimited accuracy. (In what follows we assume $\hbar = c =1$). Namely, a measuring device (or simply a test body) with zero mean velocity, having a mass
$m$ and located within the region $\delta x$ is characterized by the
gravitating energy \begin{equation}\label{graven}E = m + {\delta p^2\over m} ~,\end{equation} where $\delta p
\simeq \delta x^{-1}$. The measurement of a local characteristic of a
background metric does not allow one to take $\delta x$ very large and
therefore after minimization of Eq.(\ref{graven}) with respect to mass, $m$,
one gets an unavoidable disturbance of the background-space time. In 1959 Alden Mead showed (through a number of {\tt Gedankenexperiments}) that the combination of
Heisenberg uncertainty relations and general relativity puts absolute
limitations on the sharpness of space-time structure at the
Planck length $l_P \sim 10^{-33}$cm \cite{Mead}. It is instructive
to quote briefly his discussion regarding the
status of a fundamental length as this conceptual standpoint was
unanimous in almost all subsequent papers about space-time
uncertainties albeit many of the authors apparently did not know
this paper: \emph{"The idea is roughly as follows: Suppose there
exists a fundamental length $l$. Since a space-time coordinate
system, to be physically meaningful, must be referred to physical
bodies, it follows that no Lorentzian coordinate system can be set
up capable of specifying the coordinates of a space-time event
more precisely than $\Delta x \gtrsim l$. Conversely, if the
limitation on the coordinate system holds, the limitation on the
localizability of particles follows immediately. Thus the
fundamental length postulate may be equivalently stated as a
postulate of a limitation on realizable coordinate system. \ldots~
In terms of the light signal experiments this means, for instance,
that the time required for a light signal to propagate from body
$A$ to body $B$ and back (as measured by a clock at $A$) is
subject to uncontrollable fluctuations. However, from the point of
view of general relativity, it is completely equivalent to define
the coordinates associated with each body and clock reading by
some arbitrary convention, and to regard the light signal
experiments as yielding information about the space-time metric
associated with the coordinate system so defined. From this point
of view, fluctuations in the results of light-signal experiments
are to be regarded as indicating fluctuations in the metric, i.e.,
in the gravitational field. Thus it seems qualitatively plausible
that a fundamental length postulate is equivalent to a postulate
about gravitational field fluctuations.''} \cite{Mead}. Mead's consideration tells us
that any length undergoes fluctuations at least of the order of $\sim
l_P$. Following this reasoning of discussion another interesting observation
concerning a distance measurement for Minkowskian space-time was made by
K\'arolyh\'azy and his collaborators \cite{Karol}. With respect to the
K\'arolyh\'azy uncertainty relation the distance $t$ in Minkowski
space-time can not be known to a better accuracy than
\begin{equation} \label{KarolUncer} \delta t = \beta\,
  t_P^{2/3}\,t^{1/3}~,\end{equation} where $\beta$ is a numerical factor of
order unity . It is worth to notice that the papers \cite{Mead, Karol} sank
into oblivion for a long time\footnote{ As it is enlightened in \cite{MW}, except
for interest on the part of a few theorists who found the
discussion of \cite{Mead} convincing enough and felt that the idea
of $l_P$ as a possible fundamental length can be taken seriously,
this idea seemed to be totally unacceptable for most of
physicists. Many of the results of \cite{Mead, Karol} were
"rediscovered" in 1980s and 1990s, see for instance \cite{PNvD}.}.

Following
the discussion presented in paper \cite{CKN} one can look at
Eq.(\ref{KarolUncer}) as a relation between UV and IR scales in the framework
of an effective quantum field theory satisfying black hole entropy bound. For an effective quantum field theory in a box of size
$l$ with UV cutoff $\Lambda$ the entropy $S$ scales as, $S \sim l^3\Lambda^3$.
Nevertheless, considerations involving black holes demonstrate that the maximum entropy in a box of volume $l^3$ grows only as the
area of the box. A consistent
physical picture can be constructed by imposing a relationship between UV and
infrared (IR) cutoffs \cite{CKN}
\begin{equation}\label{BHbound}l^3 \Lambda^3 \lesssim S_{BH} \simeq \left({l\over l_P}\right)^2~,
\end{equation}
where $S_{BH}$ is the entropy of a black hole of size
$l$. Consequently one arrives at the conclusion that the length $l$, which serves as an IR cutoff, cannot be chosen
independently of the UV cutoff, and scales as $\Lambda^{-3}$. Rewriting this
relation wholly in length terms, $\delta l \equiv \Lambda^{-1}$, one arrives
at the Eq.(\ref{KarolUncer}).

\section{Minkowskian space-time in light of K\'arolyh\'azy uncertainty
  relation}

Fluctuations of the Minkowski metric described by the Eq.(\ref{KarolUncer}) are
characterized with a classical energy density (denoted hereafter by $\rho_{clasical}$) \cite{Mazia}
\begin{equation}
\rho_{classical}\label{cenerdensi} \sim {1 \over t_P^{2/3}t^{10/3}
}~.\end{equation}
The relation (\ref{KarolUncer}) together with the time-energy
uncertainty relation enables one to estimate a quantum energy density of
the metric fluctuations of Minkowski space \cite{Mazia}. With respect
to the Eq.(\ref{KarolUncer}) a length scale $t$ can be known with a maximum precision
$\delta t$ determining thereby a minimal detectable cell $\delta t^3 \simeq
t_P^2t$ over a spatial region $t^3$. In terms of the UV and IR scales discussed
above one can look at the microstructure of space-time over a length scale $t$ as consisting
with cells $\delta t^3 \simeq t_P^2t$. Such a cell
represents a minimal detectable unit of space-time over a given
length scale and if it has a finite age $t$, its existence due to time
energy uncertainty relation can not be justified with energy
smaller then $\sim t^{-1}$. Hence, having the above relation,
Eq.(\ref{KarolUncer}), one concludes that if the age of the
Minkowski space-time is $t$ then over a spatial region with linear size
$t$ (determining the maximal observable patch) there exists a
minimal cell $\delta t^3$ the energy of which due to time-energy
uncertainty relation can not be smaller than
\begin{equation} \label{cellenergy} E_{\delta t^3} \gtrsim t^{-1}~.
\end{equation} Hence, for energy density of metric
fluctuations of Minkowski space one finds
\begin{equation}\label{qenerdensi} \rho_{quantum} \sim {E_{\delta t^3}\over \delta t^3} \sim {
1 \over t_P^2 t^2}~. \end{equation} One can say the existence of
this background energy density assures maximal stability of
Minkowski space-time against the fluctuations as the
Eq.(\ref{KarolUncer}) determines maximal accuracy allowed by the
nature. Similar ideas were elaborated in \cite{Sasakura}. On the
basis of the above arguments one can go further and see that due
to K\'arolyh\'azy relation, the energy $E$ coming from the time
energy uncertainty relation $E\,t \sim 1$ is determined with the
accuracy $\delta E \sim  E \delta t/ t$. Respectively, one finds
that the energy density $\rho = E / \delta t^3$ is characterized
by the fluctuations $\delta \rho = \delta E/ \delta t^3$ giving
\begin{equation}\label{reldensfluct}{\delta \rho \over \rho}\sim
{\delta t \over t} \sim \left({t_P \over
t}\right)^{2/3}~.\end{equation}

\section{Energy budget of the universe due to K\'arolyh\'azy relation}

In the framework of inflationary cosmology the history of our
universe encompasses inflationary stage followed by the radiation
dominated and then matter dominated phases. The present
cosmological data shows that we have already left the matter
dominated phase for the dark energy dominated one which took place
only recently (at $z \approx 0.3$) \cite{Dol}. Generalization of
our approach in presence of the energy components
$\rho_{inflaton}$, $\rho_{radiation}$ and $\rho_{matter}$ is
straightforward. Due to time-energy uncertainty relation, the
energy of the cell $\delta t^3$ determined by the space-time
uncertainty relation can not be smaller than \[E_{\delta t^3}
\gtrsim  t^{-1}~,\] but now this energy contains a portion of the
inflaton or radiation + matter energy that should be subtracted
for estimating a quantum energy density of the metric
fluctuations. In what follows we will assume a minimal time-energy
uncertainty which seems reasonable as most of the history of the
universe is successfully described by the thermal equilibrium
approach. From this point of view, let us first consider the
inflationary stage during of which scale factor grows nearly
exponentially in time.

For simplicity let us
take a pure de Sitter phase as an approximation to the
inflationary stage. As it is shown in \cite{Mazia}, the
Eq.(\ref{KarolUncer}) is valid during the inflationary stage as
well if the Hubble constant during inflation $H \lesssim m_P/75^2
\approx 10^{-4}m_P$ (the timescale for the end of inflation is
taken to be $\sim 75H^{-1}$ \cite{Mukhanov}). During the inflation the energy
of a cell $\delta t^3$
contains a fraction of inflaton energy of the order of $ \sim \delta t^3H^2m_P^2 $. Hence, the energy density of the background metric
fluctuations takes the form

\begin{equation} \rho_{quantum} \simeq {1 \over t\delta t^3}-
{3\,H^2\,m_P^2\over 8\pi} ~,\end{equation} as long as
\begin{equation} {1 \over t\delta t^3}- {3\,H^2\,m_P^2\over 8\pi} \gtrsim {1
\over t_P^{2/3}t^{10/3} } ~,\end{equation} i.e. $t \lesssim
H^{-1}$ and then for $H^{-1} \lesssim t \lesssim 75H^{-1}$ follows
its classical expression, Eq.(\ref{cenerdensi}), as there is no room left any
more for the $\rho_{quantum}$ due to inflaton energy. As the energy
density of background metric fluctuations decays very fast during
the inflation it does not affect appreciably the inflationary
picture. To the end of inflation the fluctuations in the energy
density implied by the K\'arolyh\'azy uncertainty relation,
Eq.(\ref{reldensfluct}), takes the form
\[ \left.{\delta \rho \over \rho}\right|_{\,t\sim 75 H^{-1}} \sim
\left({t_P\,H \over 75}\right)^{2/3} \lesssim 10^{-5},~
\mbox{when}~ H \lesssim 10^{-6}m_P~,
\] giving thereby a constraint on the Hubble
constant during the inflation.

During the thermal history of the universe (that is, after the inflationary
stage) a requirement due to time-energy uncertainty relation that the energy of the cell $\delta t^3$ can
not be less than $t^{-1}$ can be simply summarized by the relation

\begin{eqnarray}\label{flucenbud} \left(\rho_{radiation} +
\rho_{matter} + \rho_{quantum} \right) \delta t^3 \simeq t^{-1}~.
\end{eqnarray} So we get a sort of cosmic sum rule. This
result immediately tells us that the value of $\rho_{quantum}$
depends on the fractional contribution $\rho_{radiation} +
\rho_{matter}$ to the Friedmann equation. Hence, when the energy
density due to first two terms in Eq.(\ref{flucenbud}) saturates
the cosmic sum there remains almost no room for the dark energy
$\rho_{quantum}$ and it is given by its classical expression,
Eq.(\ref{cenerdensi}), which is so small that can not be
appreciable during the cosmological evolution. For
Eq.(\ref{KarolUncer}) the relation (\ref{flucenbud}) takes the
form

\begin{equation}\label{cossumrule} \left(\rho_{radiation} + \rho_{matter} +
\rho_{quantum} \right) \beta^3\, t_P^2t^2 \simeq
1~,\end{equation} which after relating $t$ to the Hubble parameter
looks similar to the cosmic sum rule obtained from Friedmann
equation for a spatially flat metric. Namely, by taking into
account that the age of the universe during its thermal history is
$t = 1/2H,~ 2/3H$ during radiation and matter dominated phases
respectively\footnote{For simplicity we assume an instantaneous transition
  from radiation domination to the matter domination.}, the relation (\ref{cossumrule}) in light of the
Friedmann equation
\begin{equation}\label{Friedmann}H^2={8\pi t_P^2\over
3}\left(\rho_{radiation} + \rho_{matter} + \rho_{quantum}
\right)~, \end{equation} tells us that not to create changes in the expansion
rate of the universe at earlier epoches the parameter $\beta$ should satisfy
\[\beta^3 \simeq {32\pi\over 3}~, ~~~~\mbox{during the radiation domination},\]
and
\[\beta^3 \simeq {72\pi\over 12}~, ~~~~\mbox{during the matter domination}.\]
 By taking for the present
(dark energy dominated) epoch $t
\simeq H^{-1}$ from Eqs.(\ref{cossumrule},\,\ref{Friedmann}) one
finds\footnote{The lower bound on the present age of the Universe can be
  established estimating the ages of various objects it consists of. For
example, the temperature of the coldest white dwarfs in globular
clusters yields a cluster age of $12.7 \pm 0.7$Gyr \cite{Hansen}.
This gives $H_0t_0 > 0.93 \pm 0.12 $ in clear disagreement with the matter
domination where the age is estimated as $2/3H_0$.} \[\beta^3 \simeq {8\pi \over 3} ~.\] So, $\beta$ should satisfy this
relation if we want the universe to accelerate in the recent past, i.e., to get
$\rho_{quantum} \gtrsim \rho_{matter}$ for the present epoch.

Certainly, a qualitative discussion based on the combination of uncertainty
  relations with the gravity for estimating an
  uncertainty in space-time distance measurement does not allow one
 to determine (in an unique manner) the parameter $\beta$ in Eq.(\ref{KarolUncer})
with such a precision. But what seems interesting is that
numerical estimates of $\beta$ in various {\tt
Gedankenexperiments} \cite{Mazia, Mazia1} are quite close to the
above depicted values and, most important, as we have seen the
predictive consistency of the K\'arolyh\'azy uncertainty relation
with the cosmology requires a slight decay of $\beta$ during the
cosmological evolution ($\beta \simeq 3.22,\,2.66,\, 2.03$ during
radiation, matter and dark energy dominated stages respectively),
which may be attributed to the decay of radiation temperature of
the universe, as it implies the decay of corresponding thermal
fluctuations of the measuring device allowing one to perform a
space-time measurement more precisely, or to the (almost
inappreciable) short distance modification of gravity as the
K\'arolyh\'azy relation is based on the behavior of gravity at the
distance $\sim\delta t$ (see for instance \cite{Mazia}), which for
the time corresponding to the end of inflation, $t\sim 10^9t_P$,
to the radiation matter equality epoch, $t\sim 10^{32}t_P$, and to
the present epoch, $t\sim 10^{60}t_P$, gives the length scales,
$\sim 10^{-30}\mbox{cm},~10^{-22}\mbox{cm},~10^{-13}\mbox{cm}$,
respectively \footnote{Even the scale $\sim 10^{-13}\mbox{cm}$
corresponding to
  the $t\sim 10^{60}t_P$ is much smaller than the present lower experimental bound on the Newtonian inverse square law
\cite{HKH}.}.

One can say the above described approach uses a minimal setup in a
form of the basic principles of quantum mechanics and general
relativity compared to the assumptions and conjectures underlying
the basis for other approaches relating dark energy to the
(micro)stricture of space-time \cite{Causal, Calmet, Pady}. One of
the key points used in these papers is to look at the cosmological
constant as a canonically conjugate variable to the four volume of
space-time and write down for those quantities uncertainty
relation like to other Heisenberg relations. Another essential
point used by these papers is to conjecture the number of cells of
space-time (which are usually considered to have the Planck size
in contrast to what comes from the K\'arolyh\'azy relation) to fluctuate according to the Poisson
distribution. The most troublesome aspect of these approaches is
that the ever-present $\Lambda$ induced in such a way is hard to
reconcile with the early cosmology \cite{Causal, Calmet, Pady,
Barrow}.

To summarize, let us start with the prescription concerning
quantum calculus of space-time defined by the K\'arolyh\'azy
uncertainty relation \cite{Mazia}. The space-time uncertainty
relation given by Eq.(\ref{KarolUncer}) is valid for Minkowskian
space (\cite{Karol}, Ng and van~Dam \cite{PNvD}, \cite{Mazia}) as
well as for de Sitter space during the inflationary stage
\cite{Mazia} (that is, for space-time distances smaller or
comparable to the duration of inflation $\sim 75H^{-1}$
\cite{Mukhanov}). The derivation of space-time uncertainty
relation for some particular background space-time requires a
separate consideration. Space-time uncertainty relation allows one
to estimate the classical energy density of the corresponding
metric fluctuations \cite{Mazia}. On the other hand it provides a
way for estimating a quantum energy density of the metric
fluctuations with the use of time-energy uncertainty relation.
Namely, space-time uncertainty relation determines a minimal
detectable cell $\delta t^3$ over a region with linear size $t$
and if the space-time has a finite age, $t$, the energy of this
cell can be estimated by using time-energy uncertainty relation
$E_{\delta t^3}\simeq t^{-1}$ \cite{Mazia}. For estimating of
energy density associated with the metric fluctuations during the
cosmological evolution of the universe one should subtract the
contribution of the inflaton or radiation+matter energy from the
energy of $\delta t^3$ estimated through the time-energy
uncertainty relation. In this way one gets a sort of cosmic sum
rule, Eq.(\ref{flucenbud}), which exhibits that by assuming a
slight decay of $\beta$ during the cosmological evolution
(\emph{which is in the range of {\tt Gedankenexperiment}
estimates} \cite{Mazia}, \cite{Mazia1}) the K\'arolyh\'azy
uncertainty relation (\ref{KarolUncer}) can be simply reconciled
with all cosmological epoches, that is, not to disturb appreciably
the early cosmology and at the same time give a correct value for
the dark energy density. On the other hand this slight decay of
$\beta$ can be understood either as a result of temperature decay
of the radiation during the cosmological evolution which makes the
measurement procedure more precise as it implies the decay of the
corresponding thermal fluctuations of a measuring device or as a
(almost inappreciable) short distance modification of gravity
below the lengths scale $\sim 10^{-13}$cm. It should be emphasized
that the relation (\ref{KarolUncer}) with a fixed value of $\beta$
does not suffer from inconsistency as such with the cosmology if
it satisfies \[\beta^3 \simeq {32\pi\over 3}~, \] but in this case
it cannot provide sufficient amount of dark energy at present.

\vspace{0.4cm}

\begin{acknowledgments}

It is a pleasure to thank M.~Makhviladze for his help during the work on this
paper in Tbilisi and Professors Jean-Marie Fr\`{e}re and Peter Tinyakov for
invitation and hospitality at the\emph{ Service de Physique Th\'eorique,
  Universit\'e Libre de Bruxelles}, where this paper was finished. The work was
supported by the \emph{INTAS Fellowship for Young Scientists} and
the \emph{Georgian President Fellowship for Young Scientists}.

\end{acknowledgments}

\end{document}